# Highly-efficient, narrow-linewidth Brillouin microlasers implemented in compact thin-film lithium niobate microresonators

YINGNUO QIU,[1,2] CHUNTAO LI,[3,4,9] RENHONG GAO,[4] XIAOCHAO LUO,[1,2] LINGLING QIAO,[1] MIN WANG,[4] JINTIAN LIN,[1,2,*] YA CHENG[3,4,5,6,7,8,10]

[1]State Key Laboratory of Ultra-intense Laser Science and Technology, Shanghai Institute of Optics and Fine Mechanics, Chinese Academy of Sciences, Shanghai 201800, China
[2]Center of Materials Science and Optoelectronics Engineering, University of Chinese Academy of Sciences, Beijing 100049, China
[3]State Key Laboratory of Precision Spectroscopy, East China Normal University, Shanghai 200062, China
[4]The Extreme Optoelectromechanics Laboratory (XXL), School of Physics and Electronic Science, East China Normal University, Shanghai 200241, China
[5]Shanghai Research Center for Quantum Sciences, Shanghai 201315, China
[6]Hefei National Laboratory, Hefei 230088, China
[7]Collaborative Innovation Center of Extreme Optics, Shanxi University, Taiyuan 030006, China
[8]Collaborative Innovation Center of Light Manipulations and Applications, Shandong Normal University, Jinan 250358, China
[9]E-mail: 10185300303@stu.ecnu.edu.cn
[10]E-mail: ya.cheng@siom.ac.cn
*jintianlin@siom.ac.cn

**Stimulated Brillouin microlasers offer chip-scale light sources with high spectral purity and low phase noise—key attributes for applications spanning precision metrology, quantum technologies, and coherent information processing. However, simultaneously bringing both pump and scattered waves into resonance often compromises photon confinement or modal volume, resulting in limited conversion efficiency and elevated thresholds. In this work, a novel approach is proposed to generate Brillouin microlasers with high efficiency, low threshold, and narrow linewidth, by combining a cross-polarized stimulated Brillouin scattering scheme with intentional Stokes mode splitting to compensate for mode detuning. Triple-resonance and phase-matching conditions are simultaneously achieved in a 114-μm-diameter thin-film lithium niobate (TFLN) microresonator, enabling precise alignment with both the ~10 GHz Brillouin shift and the ~100-MHz narrow gain bandwidth. The resulting Brillouin microlaser achieves a narrow intrinsic linewidth of 2.88 Hz, a short-term integral linewidth of 185 Hz, an on-chip conversion efficiency of 57.92%, and a pump threshold as low as 1.03 mW. Both the conversion efficiency and the lasing threshold represent record-high performance for the TFLN platform to date.**

Brillouin microlasers, which arise from coherent photon–phonon interactions through stimulated Brillouin scattering (SBS) process to provide strong optical gain that overcomes round-trip loss in high-Q optical microresonators, have emerged as a key platform for generating on-chip narrow-linewidth laser sources. These microlasers feature a narrow gain bandwidth [1–11], a microwave-compatible gigahertz frequency shift, and a distinct acoustic dissipation mechanism. As a result, they offer high-spectral-purity light sources with small footprints [2–4,6,7,12–15], low pump consumption, and compatibility with mass-manufacturing, enabling diverse applications including coherent communications, next-generation data center networks, high precision metrology, and miniaturized atomic clocks [16,17]. Backward Brillouin microlasers have been demonstrated in a variety of material platforms, such as silicon [2], silica [5–7,9,18,19], $CaF_2$ [1], silicon nitride [4], chalcogenide [20], diamond [21], and thin-film lithium niobate (TFLN) [10,11,22], all requiring that the angular frequency difference between the pump and scattered waves matches the mechanical frequency of the medium. In silica and TFLN, this Brillouin mechanical frequency is typically around 10 GHz, whilst the gain bandwidth is only on the order of 100 MHz. Consequently, the phase matching condition of the SBS process impose stringently resonator geometric constraints in the high-Q resonators. And the most common approach to generating Brillouin microlasers has been to tailor the resonator size of millimeter or centimeter scales, ensuring that one or more free spectral ranges (FSRs) precisely match this mechanical shift. However, this method inevitably leads to large mode volumes and often results in multi-wavelength cascading Brillouin laser emission, which in turn limits the conversion efficiency, linewidth performance, and threshold characteristic.

To address this challenge, Bragg grating microstructures have been employed to induce mode splitting to generate two super-modes that match the mechanical frequency in compact isotropic microresonators [9,20]. This has been successfully realized either by incorporating Bragg gratings into silica microresonator-boundaries or by imprinting photosensitivity Bragg gratings directly into photosensitive microresonators, thereby suppressing cascading effect in the small isotropic microresonators. However, the introduction of the Bragg grating comes at the expense of optical Q factors, resulting in relatively high thresholds which are on the order of 10 mW. More recently, by implementing cross-polarized SBS in high-Q TFLN microresonators via employing the non-diagonal photoelastic tensor components of the anisotropic medium [10,11,22,23], compact Brillouin microlasers free from cascading effect have been demonstrated [11,22]. These devices achieved a threshold as low as 1.81 mW and narrow linewidth of 118 Hz. Nevertheless, due to the slight detuning from the mechanical frequency, the conversion efficiency remained limited to 19.27% [22].

In this work, we propose an alternative approach to realize Brillouin microlasers, featuring low threshold, high conversion efficiency, and narrow linewidth. Here, triple-resonance and phase-matching condition are simultaneously fulfilled in a compact TFLN microresonator, by employing a cross-polarized SBS scheme [11,22] and introducing Stokes mode splitting to mitigate mode detuning, enabling precise frequency alignment. The resulting Brillouin microlaser exhibits a narrow intrinsic linewidth as narrow as 2.88 Hz, an on-chip conversion efficiency of 57.92%, and a pump threshold as low as 1.03 mW.

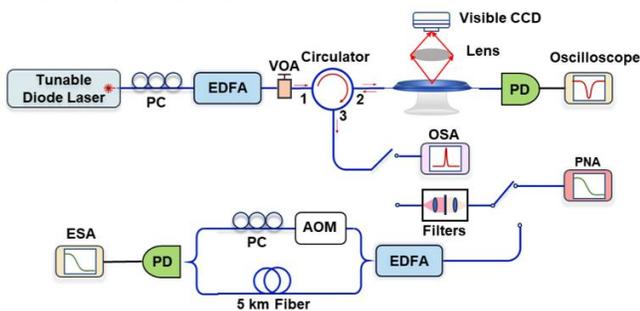

Fig. 1. Experimental setup. Here, AOM denotes acousto-optic modulator that shifts the light frequency by 100 MHz, PC is polarization controller, EDFA is erbium-doped fiber amplifier, VOA is variable optical attenuator, CCD is charge-coupled device, OSA is optical spectrum analyzer, PD is photodetector, ESA is electrical spectrum analyzer, and PNA is phase noise analyzer.

Figure 1 illustrates the experimental setup used to generate the Brillouin microlaser. A telecom-band continuous-wave (CW) tunable diode laser (model: TLB-6728, New Focus Inc.) was used as the pump source to excite the nonlinear processes in the microdisk resonator. Before entering the fiber circulator, the pump light sequentially passed through an in-line polarization controller to adjust the polarization state, an erbium-doped fiber amplifier (EDFA) for power amplification, and a variable optical attenuator (VOA) for continuous variation of the pump power. The pump light was then launched into the microdisk resonator through port 2 of the optical fiber circulator and coupled into the suspended TFLN microdisk using a tapered fiber with a waist of ~2 μm. The tapered fiber was carefully placed at the rim of the microdisk and kept in direct contact with the resonator to enable efficient light coupling into and out of the cavity. The TFLN microdisk was mounted on a six-axis motorized nano-positioning piezo-stage with a spatial resolution of 5 nm. A real-time optical microscope imaging system (composed of objective lens with a numerical aperture of 0.25 and a CCD camera) was installed above the microdisk to monitor the coupling position. To characterize the mode structure, the input laser was scanned across a spectral range of 1530 nm to 1565 nm at a low injection power of approximately 5 μW. Then, the light coupled out of the microdisk via the tapered fiber was directed to a photodetector (PD) connected to an oscilloscope, enabling real-time acquisition of the transmission spectrum during wavelength scanning. To measure the Q factors, a signal generator was used to apply an electrical triangular wave signal to the tunable laser, modulating the laser wavelength to finely scan across a resonant mode, resulting in a characteristic dip in the transmission spectrum.

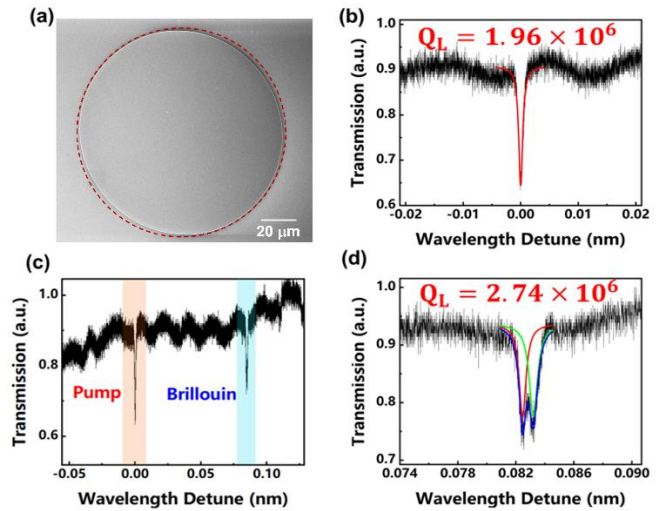

Fig. 2. (a) Scanning-electron-microscope (SEM) image of the high-Q deformed microdisk, with a red dotted line indicating a perfect circle for reference. (b) Zoomed-in view of the pump mode. (c) Transmission spectrum of the pump mode and Stokes Brillouin mode, corresponding to a wavelength difference of approximately 0.082 nm. (d) Zoomed-in view of the Stokes Brillouin mode, revealing a mode splitting.

To generate the Brillouin microlaser, the input power was increased to above the lasing threshold. The backward propagating Stokes signal generated via the SBS process was routed through the circulator and directed from port 3 to an optical spectrum analyzer (OSA) for spectral measurement with a resolution of 0.02 nm. To precisely determine the wavelength spacing between the pump and the Stokes signals, both the backward SBS signal and the backward-scattered pump signal were simultaneously fed into a high-speed photodetector (PD) connected to a real-time electrical spectrum analyzer (ESA), enabling measurement of the resulting beat-note microwave signal. To characterize the linewidth of the backward Brillouin laser signal, a tunable narrow bandpass filter with a bandwidth of 0.8 nm was used to separate the backward-scattered pump light. A home-built optical delayed self-heterodyne interferometer based on a short-delayed self-heterodyne method [24] was used to measure the phase noise (with a ~5 km single-mode fiber in the delay arm). However, as the delay fiber was much shorter than the coherence time of the Brillouin laser, resulting in a discrepancy of the measured intrinsic linewidth from the actual one. To further improve the measurement accuracy, a commercial laser phase-noise analyzer (PNA) based on the correlated self-heterodyne method [25] was also employed. This approach yields the short-term integral linewidth with a 1-ms integration time and eliminates the effect of the broadening of the Gaussian spectrum inherent to the decoherence measurement techniques like the short-delayed self-heterodyne method, thereby providing a more reliable estimate of the intrinsic linewidth.

In the experiment, a suspended TFLN microdisk resonator with a diameter of approximately 114 μm was employed to generate stimulated Brillouin lasing (SBL). This microdisk features a deliberately deformed boundary, slightly deviating from a perfect circular outline. It was fabricated by femtosecond laser photolithography-assisted chemo-mechanical etching [11,22], achieving high-Q factors. An optical microscope image of the fabricated microdisk is shown in Fig. 2(a). This slight boundary deformation introduces no significant scattering loss, but induces mode splitting in the fundamental transverse-electric ($TE_0$) modes. The measured mode profile around 1559.3 nm is presented in Fig. 2(c), showing two modes with a wavelength separation of ~0.082 nm. The mode at the shorter wavelength of 1559.264 nm (designated as the pump mode for SBS) corresponds to the fundamental transverse-magnetic ($TM_0$) mode, and exhibits a loaded Q factor of $1.96\times10^6$, as shown in Fig. 2(b). In contrast, the mode at the longer wavelength of 1559.346 nm displays a mode splitting of approximately 0.001 nm, forming two super-modes with one suitable for resonance with the Stokes wave. The loaded Q factors for these super-modes reaches $2.74\times10^6$, as depicted in Fig. 2(d). The mode splitting of the Stokes mode would provide an additional freedom of degree to mitigate the mode detuning in the SBS process, which is crucial for realizing highly efficient SBL.

The microresonator coupled with the tapered fiber was shown in Fig. 3(a). When the input pump wavelength was tuned to 1559.264 nm, a backward-scattered SBL signal was observed at a wavelength of 1559.346 nm, as shown in Fig. 3(b). The corresponding Brillouin frequency shift is approximately 0.0820 nm, corresponding to a Stokes Brillouin shift $\Omega_B$ of ~10 GHz [11,22]. And this Brillouin shift $\Omega_B$ was further confirmed by the detected radio-frequency (RF) beat note microwave signal with a finer resolution by means of optical heterodyne method using the ESA. This RF signal was centered at 10.14 GHz frequency, as shown in Fig. 3(c), corresponding to a wavelength interval of 0.0822 nm. This beat note matched the modal separation obtained in Fig. 2(c). And the Stokes light was attributed to be resonant with the longer-wavelength super-mode obtained by mode splitting. Figure 3(d) shows the on-chip Brillouin laser output power as a function of the input pump power. The measured threshold power of the Brillouin laser is approximately 1.03 mW. By linearly fitting the relationship between the output power and the pump power, the Brillouin conversion efficiency is estimated to be 57.92%. Both these results surpass the records reported on the TFLN platform so far [10,11,22].

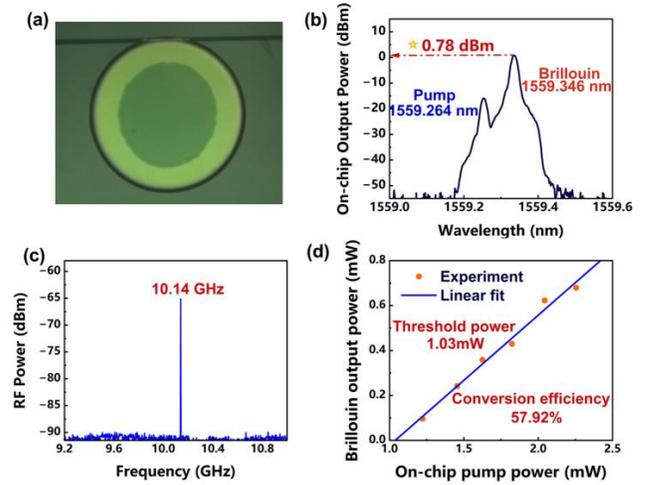

Fig. 3. (a) Optical microscope image of the microresonator coupled with the tapered fiber. (b) Optical spectrum of the SBL signal. (c) Microwave beat note between the pump wave and the SBL signal. (d) Output power of the SBL as a function of on-chip pump power.

To qualitatively understand the high conversion efficiency and the low threshold observed in the experiment, the system dynamics can be described using the coupled-mode equations under the slowly varying envelope approximation [26]. In typical Brillouin laser systems, the mechanical damping rate usually satisfies $\gamma_m \gg \kappa_s$ ($\kappa_s$ is the decay rate of the optical scattering wave in the microresonator), such that the mechanical mode can be considered to respond instantaneously. Under this approximation, the Brillouin gain coefficient can be obtained as

$$G_B(\Omega) = \frac{4|g_0|^2}{\gamma_m}\frac{(\gamma_m/2)^2}{(\gamma_m/2)^2+(\Omega-\Omega_m)^2}. \quad (1)$$

where $\Delta_m = \Omega - \Omega_m$ denotes the detuning between the acoustic driving frequency $\Omega$ and the mechanical resonance frequency $\Omega_m$, and $g_0$ is the single-photon optomechanical coupling strength. According to Eq. (1), the Brillouin gain increases as the detuning between the acoustic driving frequency and the mechanical resonance frequency decreases. Consequently, a higher Brillouin conversion efficiency can be achieved when this detuning is small. Moreover, under the steady-state conditions, the Brillouin lasing threshold power can be expressed as

$$P_{\text{th}} \propto \Delta_p^2. \quad (2)$$

Therefore, as the detuning $\Delta_p$ between the pump laser and the cavity mode increases, the intracavity optical power decreases, leading to an increase in the Brillouin lasing threshold. Therefore, these theoretical results agree well with our experimental observations, where a high conversion efficiency and a low threshold were obtained by optimizing the triple-resonance condition.

The frequency noise and linewidth characteristic of the generated stimulated Brillouin laser were also investigated. The frequency noise spectrum measured using the short-delayed self-heterodyne method is shown in Fig. 4(a). From the white frequency noise floor of $N_{wfn}$ = 0.916 Hz$^2$/Hz, the intrinsic linewidth $L_{IL}$ of the SBL is estimated to be 2.88 Hz ($L_{IL}=N_{wfn} \times \pi$). Furthermore, the correlated self-heterodyne method was employed to characterize the frequency noise. The white frequency noise floor of the backward SBL is measured to be 58.97 Hz$^2$/Hz, corresponding to a short-term integral linewidth of 185.16 Hz, as depicted in Fig. 4(c). In comparison, the short-term linewidth of the pump laser is measured to be 313.81 Hz, as shown in Fig. 4(b). These results indicate that the obtained SBL exhibits a significant linewidth narrowing compared with the pump laser.

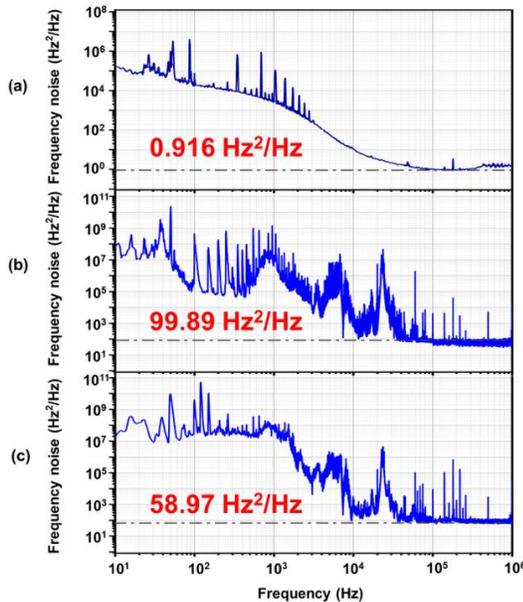

Fig. 4. (a) Frequency noise spectrum of the SBL signal measured by the short-delayed self-heterodyne method, showing an intrinsic linewidth of 2.88 Hz. Frequency noise spectra of (b) the pump and (c) the SBL signals measured by the correlated delayed self-heterodyne method, showing short-term integral linewidths of 314 Hz and 185 Hz, respectively.

In summary, an on-chip stimulated Brillouin microlaser was demonstrated in a compact lithium niobate microdisk, simultaneously achieving a high conversion efficiency, a low threshold, and low noise performance, by leveraging the combination of the cross-polarized SBS scheme and the Stokes mode splitting to mitigate mode detuning without compromising the optical Q factor, enabling precise mode alignment in the compact device, and yielding the on-chip high-performance Brillouin microlasers. These results underscore the potential of TFLN-based Brillouin photonics for realizing high-coherence light sources and compact microwave photonic systems with low loss [27-31].